\documentclass[dvipdfmx]{akari2017}

\usepackage[utf8]{inputenc}
\usepackage{graphicx}
\shorttitle{Searching for previously unknown classes of objects in AKARI-NEP Deep}
\shortauthors{A. Poliszczuk et al.}

\begin{document}

\title{Searching for previously unknown classes of objects\\
in the AKARI-NEP Deep data with fuzzy logic SVM algorithm} 



\correspondingauthor{Agnieszka Pollo}
\email{agnieszka.pollo@ncbj.gov.pl}

\author{Artem Poliszczuk} 
\affiliation{National Centre for Nuclear Research, ul. Ho\.za 69, 00-681, Warsaw, Poland} 
\author{Aleksandra Solarz} 
\affiliation{National Centre for Nuclear Research, ul. Ho\.za 69, 00-681, Warsaw, Poland} 
\author{Agnieszka Pollo} 
\affiliation{National Centre for Nuclear Research, ul. Ho\.za 69, 00-681, Warsaw, Poland}
\affiliation{The Astronomical Observatory of the Jagiellonian University, ul. Orla 171, 30-244, Krak\'ow, Poland}
\author{the NEP-Deep Team}



\begin{abstract}

In this proceedings application of a fuzzy Support Vector Machine (FSVM) learning algorithm, to classify mid-infrared (MIR) sources from the AKARI NEP Deep field into three classes: stars, galaxies and AGNs, is presented. FSVM is an improved version of the classical SVM algorithm, incorporating measurement errors into the classification process; this is the first successful application of this algorithm in the astronomy.

We created reliable catalogues of galaxies, stars and AGNs consisting of objects with MIR measurements, some of them with no optical counterparts. Some examples of identified objects are shown, among them O-rich and C-rich AGB stars.


\end{abstract}



\setcounter{page}{1}



\section{Introduction} 

Machine learning algorithms are becoming one of the main tools in the modern, big data driven astronomy. The support vector machine (SVM) algorithm is one of the most important and efficient among them. However, the classical version of the SVM, which was applied in numerous astronomical tasks, is unable to incorporate measurement uncertainties into the classification process.

The present work is the first attempt of using measurement errors in the SVM-based classification in astronomy. Moreover, the training sample is based on spectroscopic data previously unused in automatic classification of the AKARI-NEP Deep data. As a consequence, new types of objects were found, and a more reliable source catalogue was created. In this work no optical auxiliary data were used to preserve the completeness of the IR catalogue.

\section{Data}

The AKARI NEP-Deep sky survey covers an area of 0.4 deg$^2$ around the north ecliptic pole \citep{Matsuhara}. Obtained data were collected by the AKARI Infra-red Camera (IRC) instrument, which observed sky in the nine passbands in the near- and mid-infrared: 3 $\mu$m (N3), 4 $\mu$m (N4), 7 $\mu$m (S7), 9 $\mu$m (S7), 11 $\mu$m (S11), 15 $\mu$m (L15), 18 $\mu$m (L18) and 24 $\mu$m (L24). In total it contains almost 27 thousand sources, with 23,325 sources observed in N2; 26,180 sources in N3; 26,332 sources in N4; 8,650 sources in S7; 8,516 sources in S9; 8,769 sources in S11; 10,611 sources in L15; 10,782 sources in L18 and 5,704 sources in L24. Thanks to many years of work of the AKARI NEP team members, a rich database of multivawelength observations of the NEP field is now gathered. Nevertheless, the identity of many objects remains unknown and creation of the complete catalogues of IR-selected sources in the different IRC bands remains a challenge.

In the present work a catalogue of 1,808 objects is used. Such a limitation of the number of objects was caused by the requirement of construction of the parameter space for a machine learning algorithm described in the next section. As the parameters, differences between fluxes (i.e. colours) were used: N2-N3, N3-N4, S7-S11 and L15-L18. Because of that, catalogue was made only of those objects, which were measured in all passbands used: N2, N3, N4, S7, S11, L15, L18.

\section{The SVM algorithm}

Classification of objects is usually based on their properties of some kind. These properties can be parametrized to create the parameter space. However, in most cases, separation of objects in such an input parameter space is very difficult or even impossible. The SVM is a supervised learning algorithm, what means that it needs data with a priori known labels to perform a training process and to create a classifier to be used on a sample of objects with unknown assignment. The main idea of the SVM is based on the mapping of the data from its normal (input) parameter space to the high dimensional feature space (these new dimensions are created as combinations of the input parameters), where creation of a separating hyperplane becomes possible (see Fig. 1).

\begin{figure}[!h]
	\epsscale{0.55}
	\plotone{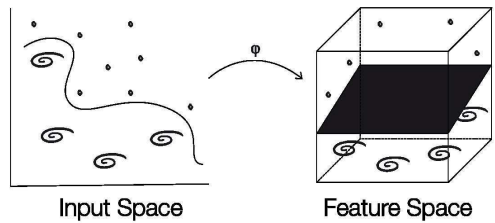}
\caption{Main idea of the SVM: mapping data from the input parameter space to the high dimensional feature space to construct a separating hyperplane
}\label{fig:header}
\end{figure}

Construction of such a hyperplane is based on the maximization of the distance between classes in the feature space.

Despite classical SVM's high efficiency (its application to AKARI-NEP Deep data is shown in \citet{Solarz,Solarz2}, it has a weakness, which becomes important in the case of application to more complex samples - it treats equally all the data. In such a situation the precision of measurement has no influence on the classification. To overcome this problem, a fuzzy SVM \citep[FSVM;][]{LinWang} was used in the present work. FSVM is a modified version of a normal SVM algorithm, which allows to tie in objects with their so called fuzzy memberships. These memberships define the confidence of the object belongingness to a particular class and determine how important a certain object would be in the classification process. Basing the fuzzy memberships on the measurement uncertainties results in an increase of the classifier's flexibility and its more realistic outcome.

The implementation of the FSVM algorithm applied in this work is able to perform an n-class classification and its performance in the case of imbalanced data sets was shown in the tests to be better than that of the classical SVM. In the case of the NEP-Deep data, 3-class task for stars, galaxies and AGNs was performed. Due to lack of the sufficient number of the spectroscopic data from the NEP-Deep field, the training sample was constructed by crossmatching the AKARI-NEP Wide data with spectroscopic observations of this region \citep{Shim}. In this way a training sample of 71 stars, 439 galaxies and 122 AGNs was created. The input parameter space was constructed of 4 colours (N2-N3, N3-N4, S7-S11 and L15-L18). Two exemplary colour-colour plots of the training data are shown in Fig. 2. One can clearly see that a simple linear separation of these three classes is not possible.

Due to imbalanced data sets, a tool for the evaluation of the classifier's performance in the training process needs to be chosen carefully. Classical evaluation metrics, such as accuracy, have a tendency to privilege bigger classes which affects the classification. Because of that a Cohen's kappa statistics \citep{Cohen} was used as an evaluation metric in the training process. This metric ($\kappa$) remains sensible to the changes of the smaller classes and shows more realistic grading of the performance of the classifier. It can take values between -1 and 1, where 1 corresponds to the best possible performance, while 0 corresponds to the random classification. The FSVM applied to the NEP-Deep data received $\kappa = 0.719$ in the training process.


\begin{figure*}[!h]
\centering
\gridline{\fig{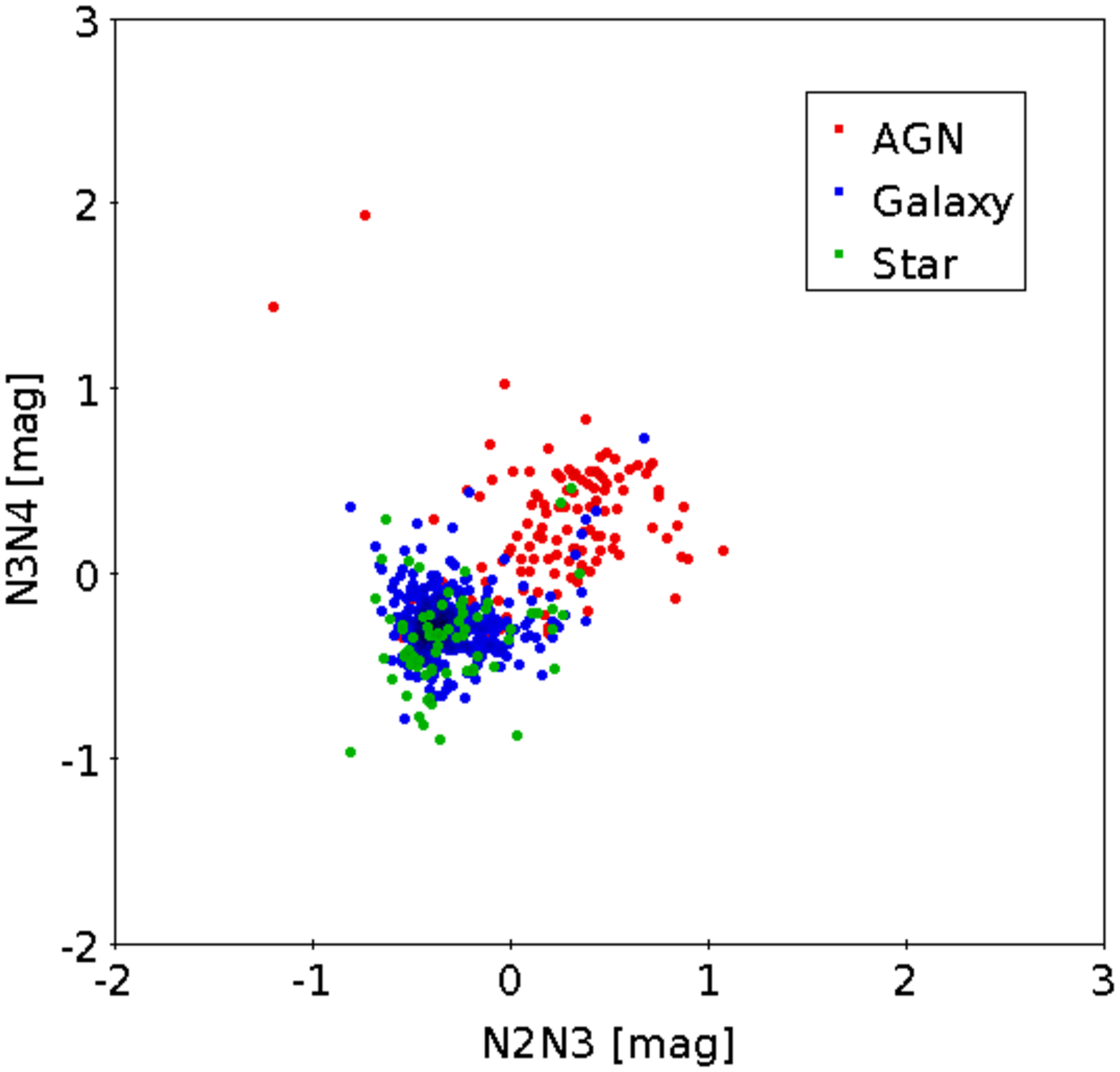}{0.36\textwidth}{}
          \fig{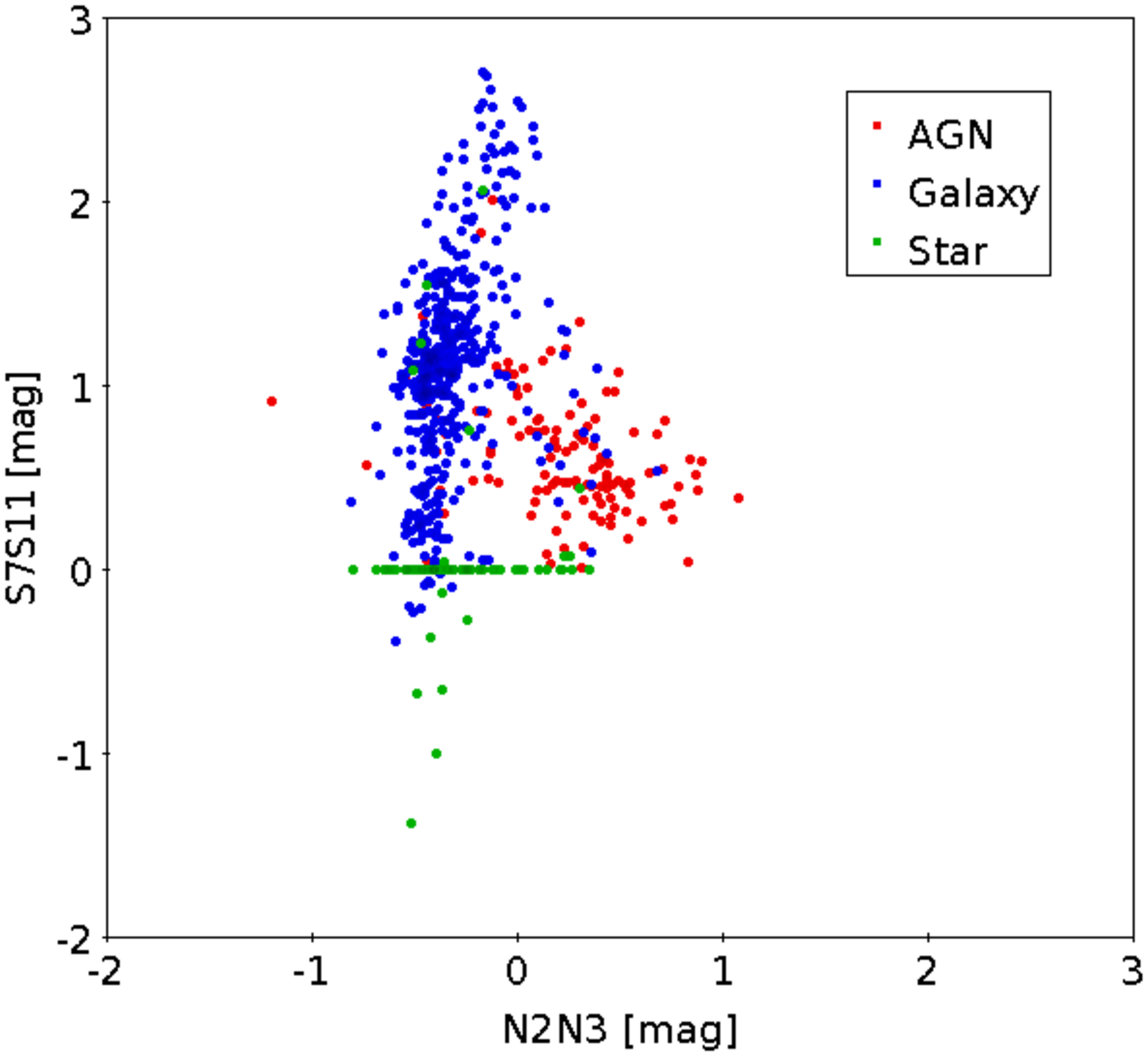}{0.38\textwidth}{}
          }
\caption{N2-N3 vs N3-N4 and N2-N3 vs S7-S11 colour-colour plots for the NEP-Deep training sample}
\end{figure*}

\section{New catalogues of the AKARI-NEP Deep data}

Generalization was performed on the sample of 1722 objects with an unknown class. As a result, a new catalogue of FSVM-classified 230 stars, 705 galaxies and 873 AGNs was obtained (exemplary color-color plots for these objects are presented in Fig. 3). Because of the lack of a complete, representative sample of stars, some of the galaxies fall into the star region (it can be seen in the bottom left corner of the N2-N3 vs S7-S11 colour-colour plot in Fig. 3).

To probe the classification's correctness, one can draw differential number counts for all the catalogues. In such counts sources per flux interval are toted up and normalized to the counts expected in the Euclidean space for a non-evolving uniformly distributed population of sources. Number counts presented in Fig. 4 display a shape expected for particular classes of objects and are consistent with previous works \citep{Murata, Arendt}. Stars classified as galaxies described in the first paragraph of the section were excluded from the number counts. One of arguments for their membership to the stellar class is their placement in the number counts, typical for the stars. Another argument is their high stellarity parameter as given by SExtractor \citep{Bertin} run on the N2 NEP images.

\begin{figure*}[!h]
\centering
\gridline{\fig{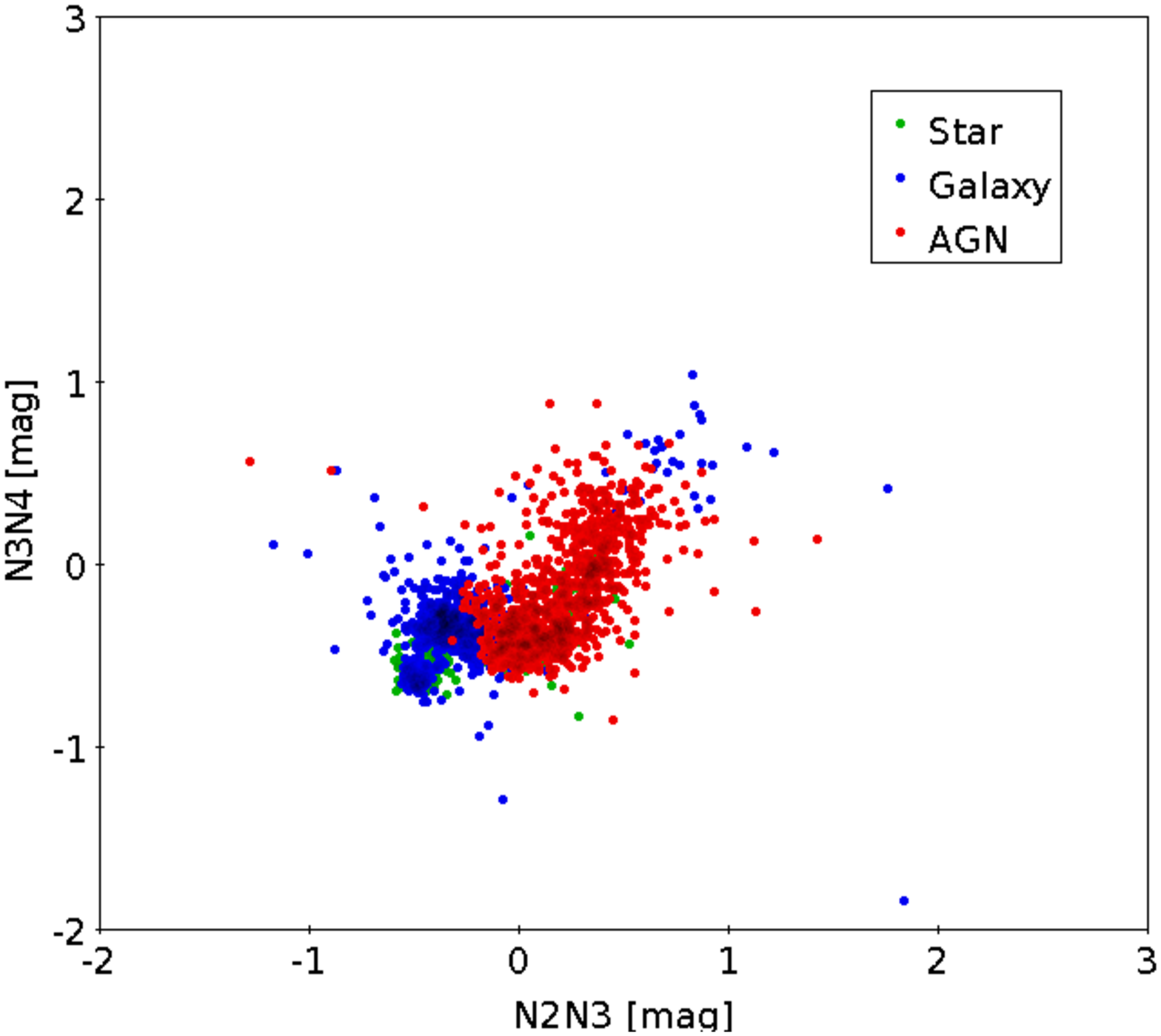}{0.36\textwidth}{}
          \fig{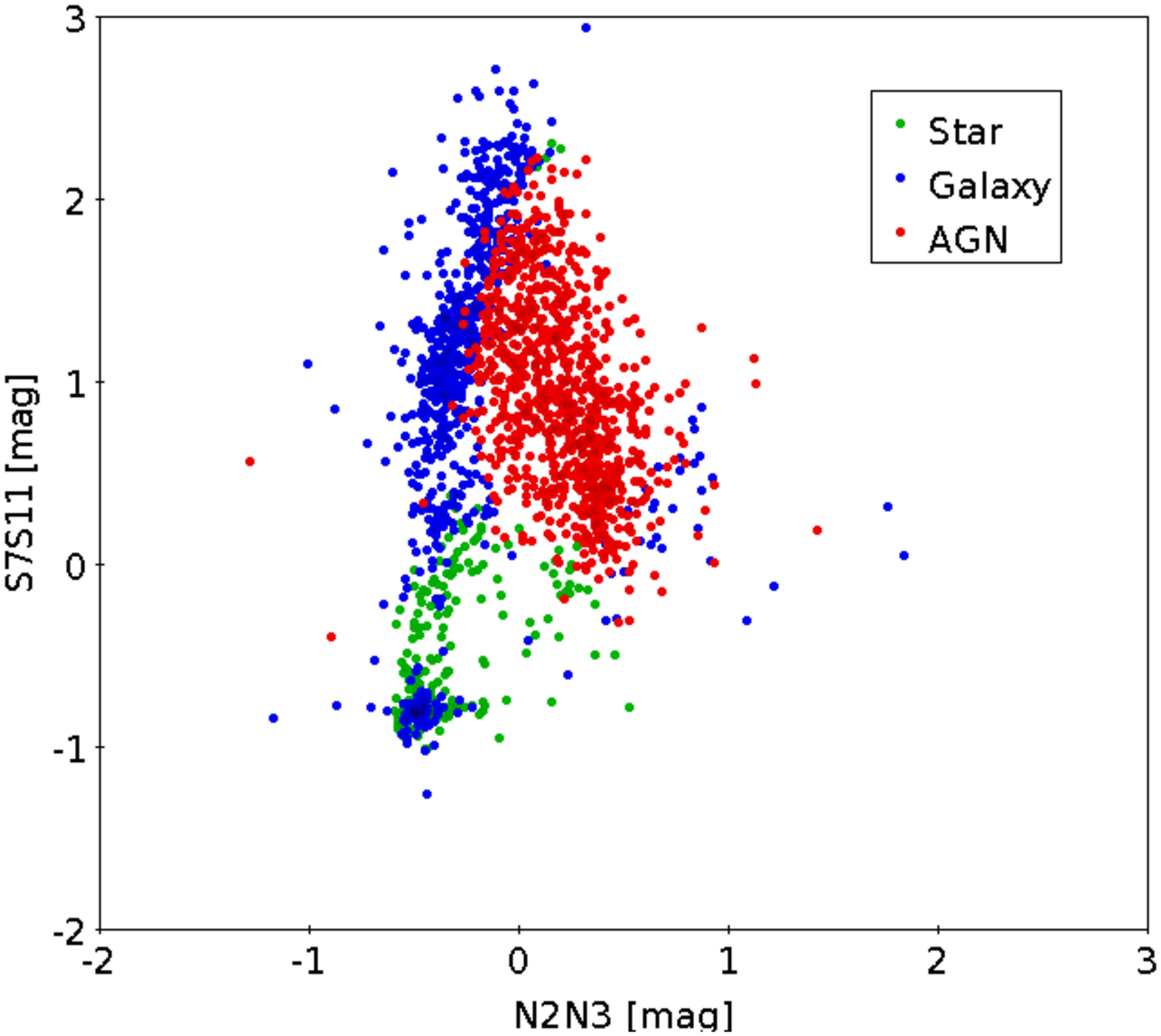}{0.36\textwidth}{}
          }
\caption{N2-N3 vs N3-N4 and N2-N3 vs S7-S11 colour-colour plots for the NEP-Deep general sample}
\end{figure*}

\begin{figure*}[!h]
\centering
\gridline{\fig{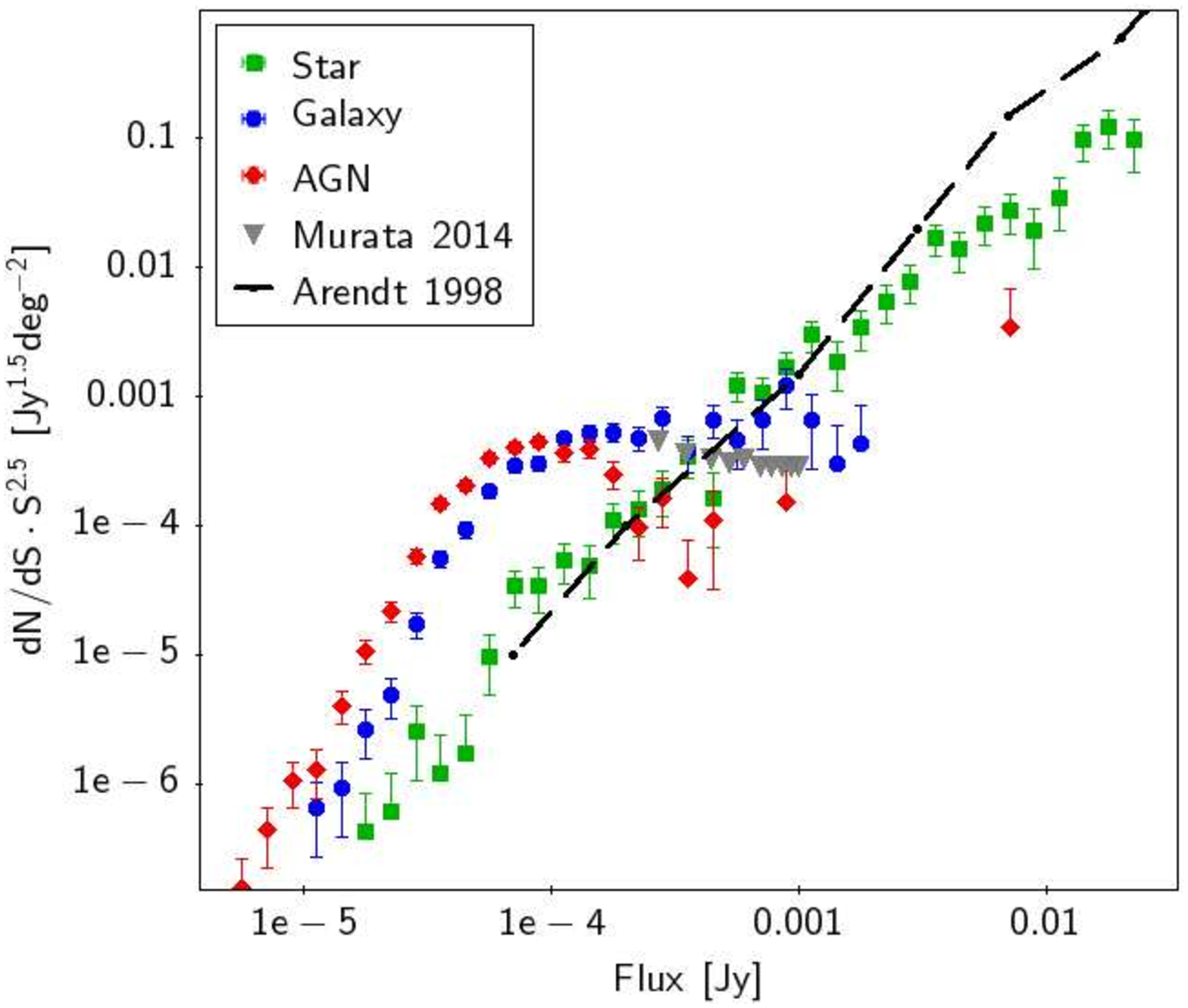}{0.4\textwidth}{}
          \fig{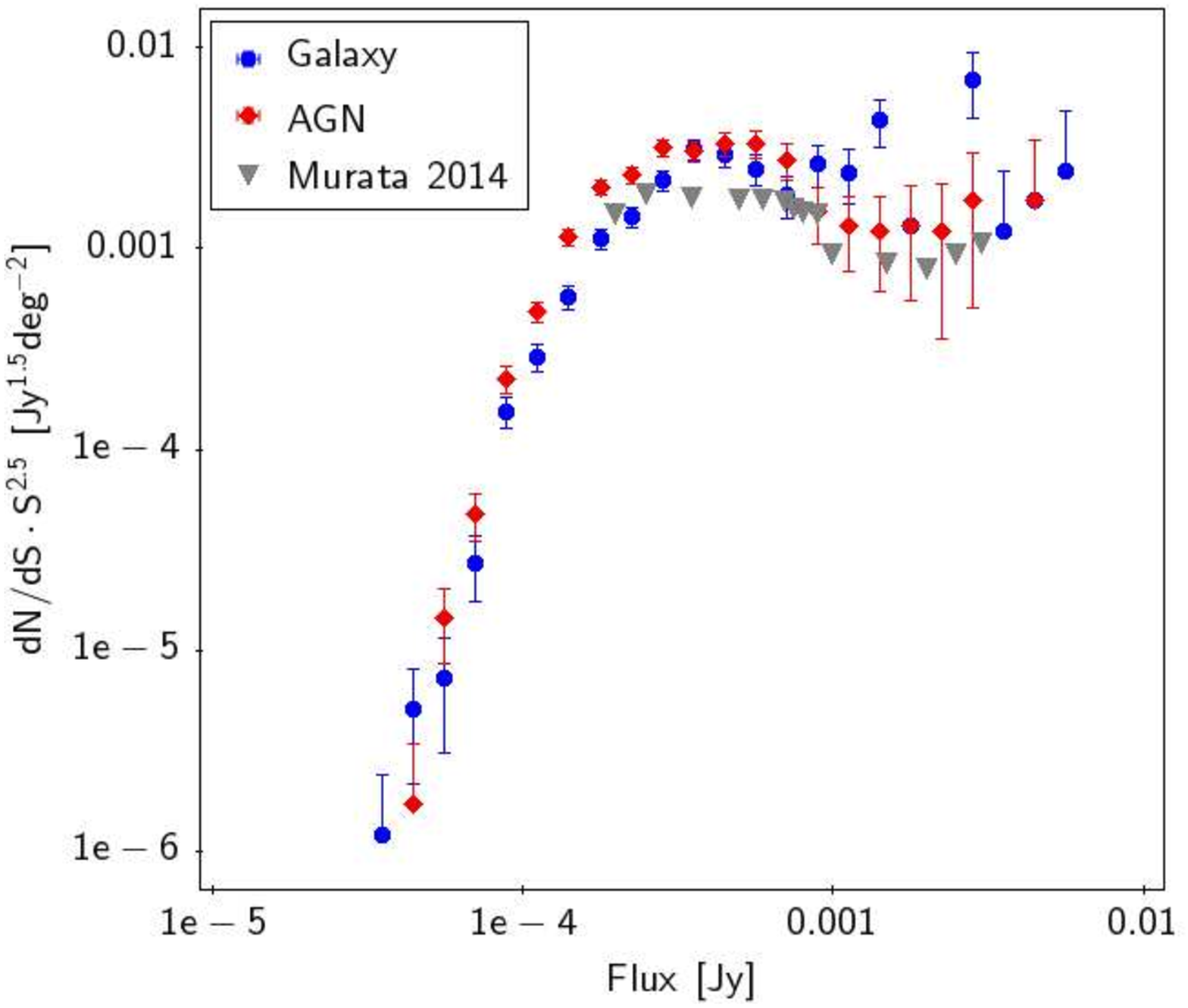}{0.4\textwidth}{}
          }
\caption{N2 and L18 number counts for the NEP-Deep general sample}
\end{figure*}


After the final catalogue was constructed, verification of all obtained classes of sources was performed. In particular, spectral energy distributions (SEDs) of objects classified as stars were fitted by the VOSA software \citep{Bayo}. Based on the effective temperature distribution, two main subclasses of stars can be found in the NEP Deep field. The first class consists of cold K, M stars and Asymptotic Giant Branch (AGB) stars. The second class includes hotter A and B type stars. Additionally, one white dwarf was found (see Fig. 5).

\begin{figure}[!ht]
	\epsscale{0.6}
	\plotone{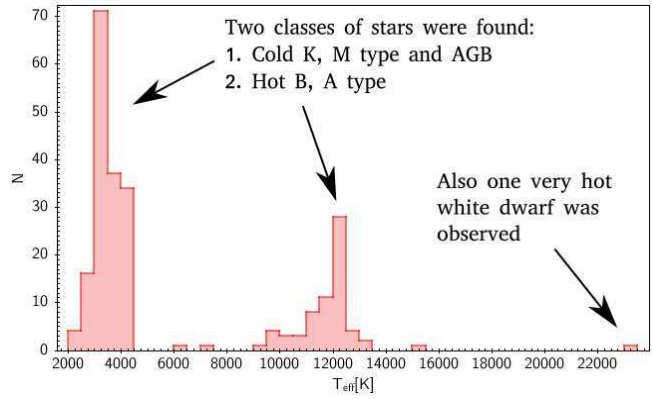}
\caption{Effective temperature distribution of stars provided by VOSA software shows two main subclasses
}\label{fig:header}
\end{figure}

In total, 127 main sequence stars and 66 AGB C-rich and 37 O-rich stars were found (MIR-to-NIR fitted SED of two exemplary O-rich and C-rich AGB stars are shown in Fig. 6).

\begin{figure*}[!ht]
\centering
\gridline{\fig{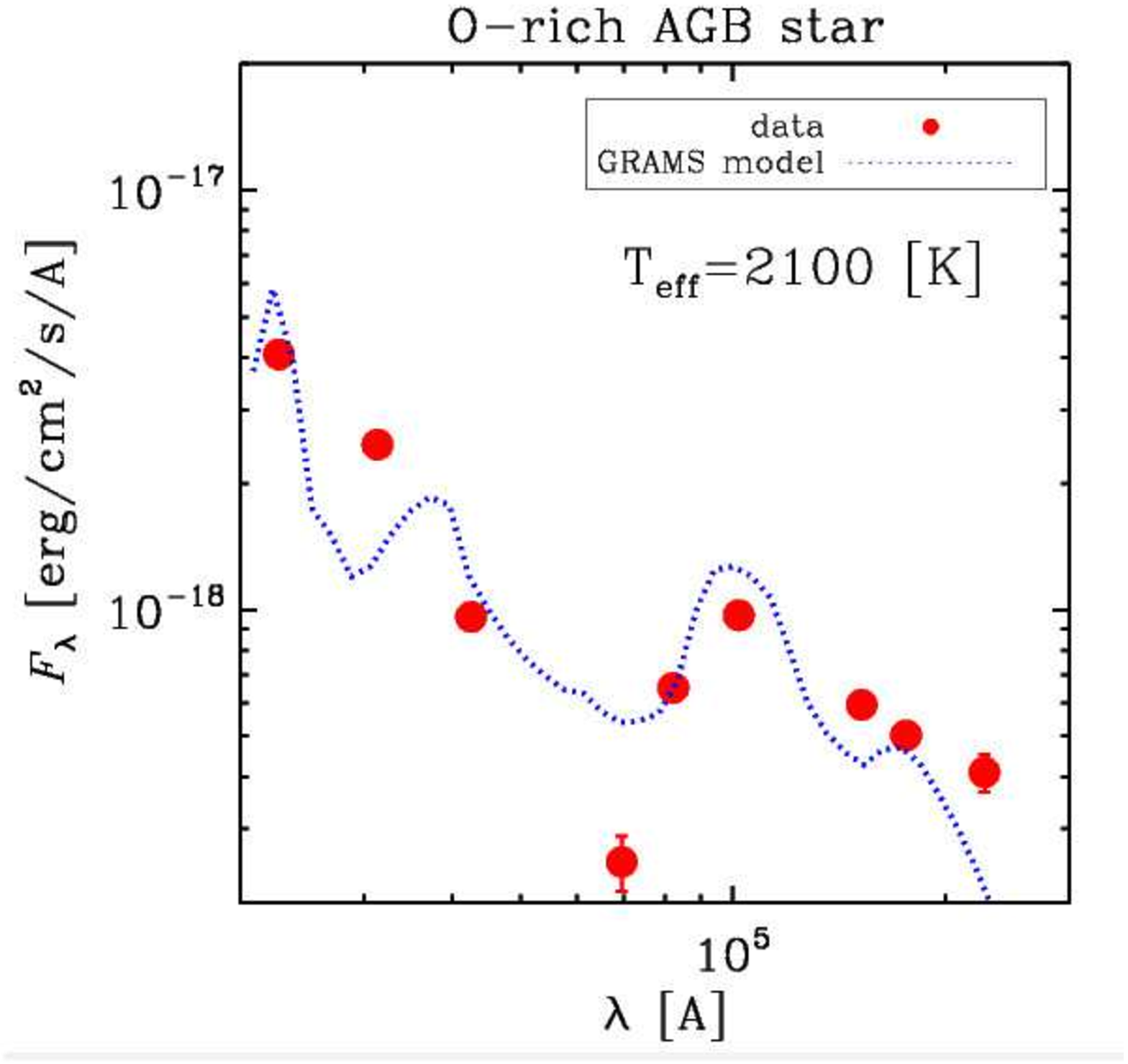}{0.36\textwidth}{}
          \fig{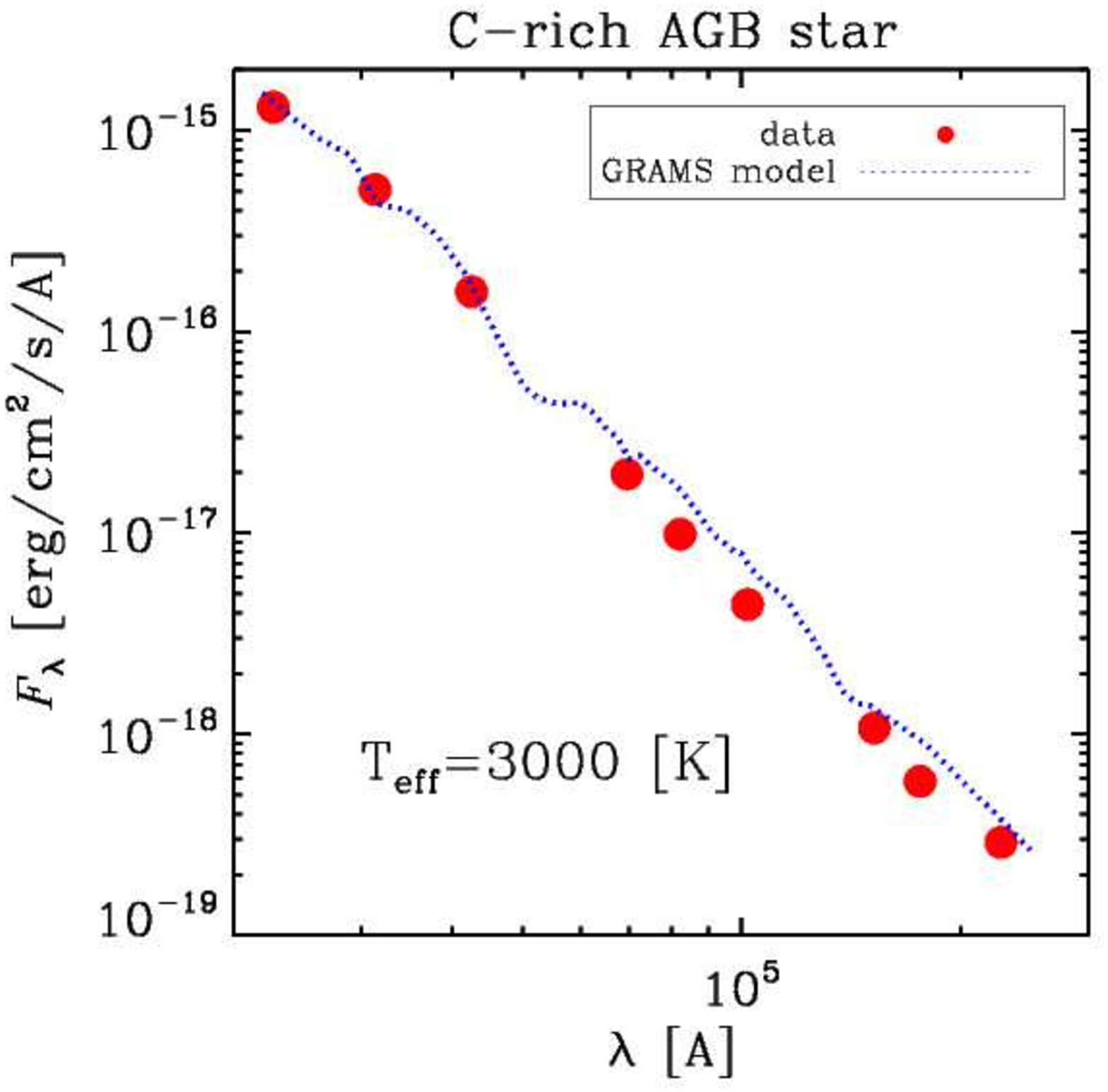}{0.36\textwidth}{}
          }
\caption{Examples of O-rich and C-rich AGB stars}
\end{figure*}

\section{Summary and Future Work}

In the present work, a new, more trustable AKARI-NEP Deep catalogue of FSVM-classified sources with a full NIR to MIR spectral coverage is presented. Presented results can be treated as an argument for using the FSVM as a more physically motivated tool than the classical SVM algorithm. However, to perform a precise analysis of the objects more complete training samples are needed and additional spectroscopic measurements must be gathered. A more detailed and extended version of the presented research will be published \citet{Poliszczuk}.

\section{Acknowledgements}

This research is based on observations with AKARI, a JAXA project with the participation of ESA. This research has been supported by National Science Centre grants number UMO-2015/16/S/ST9/00438.

\end{document}